\documentclass[galaxies,review,accept,oneauthor,10pt,a4paper]{mdpi} 

\firstpage{1} 
\articlenumber{x}
\doinum{10.3390/------}
\pubvolume{xx}
\pubyear{2016}
\copyrightyear{2016}
\externaleditor{Academic Editor: name}
\history{Received: date; Accepted: date; Published: date}
 \theoremstyle{mdpi}
 \newcounter{thm}
 \newcounter{ex}
 \newcounter{re}
 \theoremstyle{mdpidefinition}

\Title{Exoplanets: past, present, and future}

\Author{Chien-Hsiu Lee $^{1}$}
\AuthorNames{Chien-Hsiu Lee}

\address{%
$^{1}$ \quad Subaru Telescope, NAOJ; leech@naoj.org}
\abstract{Our understanding of extra-solar planet systems is highly driven by advances in observations in the past decade. Thanks to high precision spectrograph, we are able to reveal unseen companions to stars with the radial velocity method. High precision photometry from the space, especially with the Kepler mission, enables us to detect planets when they transit their stars and dim the stellar light by merely one percent or smaller. Ultra wide-field, high cadence, continuous monitoring of the Galactic bulge from different sites around the southern hemisphere provides us the opportunity to observe microlensing effects caused by planetary systems from the solar neighborhood, all the way to the Milky Way center. The exquisite AO imaging from ground-based large telescopes, coupled with high-contrast coronagraph, captured the photons directly emitted by planets around other stars. In this article, I present a concise review of the extra-solar planet discoveries, discussing the strengths and weaknesses of the major planetary detection methods, providing an overview of our current understanding of planetary formation and evolution given the tremendous observations delivered by various methods, as well as on-going and planned observation endeavors to provide a clear picture of extra-solar planetary systems.}

\keyword{Planetary systems; planets and satellites: atmosphere; protoplanetary disks; techniques: radial velocities; techniques: high angular resolution; gravitational lensing: micro }

\begin{document}

%%%%%%%%%%%%%%%%%%%%%%%%%%%%%%%%%%%%%%%%%%
%% Sections that are not mandatory are listed as such. The section titles given are for Articles. Review papers and other article types have a more flexible structure. 

%% Only for the journal Gels: Please place the Experimental Section after the Conclusions

%%%%%%%%%%%%%%%%%%%%%%%%%%%%%%%%%%%%%%%%%%
%\setcounter{section}{-1} %% Remove this when starting to work on the template.
\section{A brief history}
Searching for and characterization of planets beyond the solar system has been a long quest for observational astronomy. Interestingly, the very first extra-solar planets (hereafter exoplanets) were not found around a main-sequence star, but rather an neutron star using the change in pulsar timing \citep{1992Natur.355..145W}. The first exoplanet around solar like stars -- 51 Pegasi b -- was found by radial velocity method by Mayor and Queloz in 1995 \citep{1995Natur.378..355M}. The idea is, if there is an unseen planetary mass companion to the host star, they will both exhibit orbital motion around the common mass center, rendering periodic blue and redshift of emission and/or absorption lines in the host star's spectra due to Doppler effect. While the result received vast amount of critics, especially with the speculations that such kind of radial velocity signal could be mimicked by stellar spots, it has turned out that the radial velocity signal is truly coming from a planetary mass companion. The downside of the radial velocity method is that we do not know the inclination of the planetary orbit, so the interpreted planetary mass is always the lowest possible value, and the true mass can be much heavier if we are observing the planetary system in a almost face-on view. As can be imagined, to reveal the small perturbation caused by planetary companion in the host star's radial velocity, it requires very precise measurement of radial velocity, which is only possible via very high resolution spectrograph. Indeed, discoveries of the vast majority of exoplanets via the radial velocity method is only possible with the High Accuracy Radial velocity Planet Searcher (HARPS), a high precision echelle spectrograph build by Mayor et al. \cite{2003Msngr.114...20M} mounted on the ESO 3.6m telescope in La Silla in Chile. A similar instrument, HARPS-N, has been implemented and installed on the 3.6-m Italian Telescopio Nazionale Galileo at La Palma, Canary Islands in Spain in 2012 \cite{2012SPIE.8446E..1VC}, allowing astronomers to search for exoplanets with the radial velocity method from both hemispheres.

The next big step of exoplanet observation came in 1999, where Charbonneau et al. \cite{2000ApJ...529L..45C} detected the first transiting signal -- an exoplanet pass in front of the host star from our light of sight -- in the radial velocity confirmed exoplanet system HD 209458. The transit method provides great leverage to the study of exoplanet atmosphere. This is because during transit, the light from the host star will pass through the atmosphere of the exoplanet, thus revealing the molecular absorption features in the exosphere. By comparing the host star’s spectra during and without transit, we will be able to probe the exosphere with high S/N. In addition, variations of exospheric absorptions features among different transit cycles can also shed lights on the weather of the exoplanets. The first exoplanetary transmission spectroscopy was carried out by Charbonneau et al. in 2002 \cite{2002ApJ...568..377C}, who detected sodium absorption features in the atmosphere of a hot Jupiter HD 209458b. Subsequent works have discovered oxygen and carbon molecules in HD 209458b, e.g. Vidal-Madjar et al. in 2004 \cite{2004ApJ...604L..69V}, as well as molecules in other transiting exoplanets, e.g. HD 189733b by Swain et al. in 2008 \cite{2008Natur.452..329S} and XO-1b by Tinetti et al. in 2010 \cite{2010ApJ...712L.139T}. The only disadvantage of transit method is as the exoplanet is relatively small compared to the host star, the dimming of host star is so tiny that it requires very high photometric accuracy, at least 1\% level, to detect exoplanet via this method. Such kind of photometric accuracy is difficult to achieve from the ground, where atmospheric turbulence is of concern and prevents us to reach high accuracy photometry for faint stars and/or to search for smaller exoplanets. The transit search from ground-based telescopes have thus been limited to bright stars instead. This limitation, however, is not an issue from space-based missions. The Kepler spacecraft, for example, have enabled high precision, continuous monitoring of stars from the space, delivering more than 2,000 exoplanets via transit method\footnote{https://www.nasa.gov/kepler/discoveries}. While Kepler spacecraft faced some technical issues when two out of the four gyroscopes were out of service in the middle of 2013, scientists were able to come up with a plan to use the solar pressure as the additional "third" gyroscopes, along with the two that were still functioning, and began the Kepler second light observations (dubbed as "K2") from late 2013 and delivering high quality science data till now.

Around the time where HD209458b has been observed with the transit method, the first exoplanetary microlensing event has also been observed \citep{2004ApJ...606L.155B}. According to Einstein's general relativity, massive objects in the foreground will induce strong space-time curvature, bending lights from background sources and focusing their lights gravitaionally. If the foreground objects are as massive as galaxies or clusters of galaxies, the space-time curvature is so strong that the background sources will be lensed into multiply images separated by several seconds of arcs, which is often called strong lensing. If the foreground objects are of smaller masses, like the stars or planets, then the multiplied images will have separations of merely several micro arcseconds, hence bearing the name microlensing. Although we cannot resolve the multiplied images in the microlensing regime, we can, however, see the brightness variation of the background sources. Bohdan Paczynski was the first one to conceive the idea of using microlensing to search for unseen, compact stellar objects \cite{1986ApJ...304....1P}. While Paczynski's original proposal was to search for MACHOs, he and his (then) postdoc, Shude Mao, extend the idea and demonstrate that microlensing is capable of detecting binary stars and stars associated exoplanets \cite{1991ApJ...374L..37M}. Although Einstein has speculated that observing microlensing effects from stars is unlikely in his Science article in 1936 \cite{1936Sci....84..506E}, Paczynski has calculated probability of microlensing by stars, and it turns out that the chance (or the optical depth) is one out of a million. This suggests that if we can find a dense stellar field with more than 1 million stars serving as background sources, we will be able to observe the microlensing effects. Motivated by Paczynski's seminar paper, and with the advent of the modern camera based on CCDs, numerous observation campaigns, e.g. MACHO, EROS, OGLE, MOA, to name a few, have started continuous monitoring of dense stellar fields such as the Galactic bulge and the Magellanic Clouds. While the first microlensing events by MACHOs (or stars) were reported in 1993 by MACHO \cite{1993Natur.365..621A}, EROS \cite{1993Natur.365..623A}, and OGLE \cite{1993AcA....43..289U} callaboration, exoplanetary microlensing events were not detected until 2003 by Bond et al. \cite{2004ApJ...606L.155B}. This is because while the microlensing signals by stellar objects usually last for months, the microlensing signature by exoplanets only last for days, and it requires dedicated, very high cadence, multi-site follow-up to really distinguish and pin-down the exoplanetary microlensing signature from other effects, e.g. binary stars, parallax, and so forth. 

The only disadvantage of microlensing is that, unlike other methods, the exoplanetary signal will not repeat, and we have only one chance to gasp the planetary nature from microlensing light curves. Nevertheless, it has been shown that given well-sampled light curves, we can indeed extract most of the fundamental planetary parameters, such as the mass and orbital distance without ambiguity. This is especially if we have simultaneous observations from the ground and from space (e.g. Spitzer), which will provide the parallax of the lens and further constrain the lens parameter \citep{{2015ApJ...799..237U}}. In addition, the benefit of microlensing is that it only relies on the gravitational signal of the planet, but does not utilize the lights from the host star nor from the exoplanet itself, we can therefore detect very low mass (e.g. earth-like) exoplanet out to a very large orbits (beyond the snow-line), which is otherwise not possible with radial velocity, transit, or direct imaging methods. Furthermore, exoplanets could be found by microlensing and/or pixel lensing in other galaxies as well, e.g. in M31 by Ingrosso et al. \citep{2009MNRAS.399..219I}, or by new attempts to explain quasar light curve as the intervening of planet-like objects \citep{2018ApJ...853L..27D}.

In 2008, astronomers were able to detect photons directly emitted from the exoplanets for the first time. This is only possible with the exquisite spatial resolution enabled by ground-based AO instrument, coupled with high contrast imaging capability, delivered by the Keck and Gemini telescopes at that time. With these cutting edge technology, Marois et al. \cite{2008Sci...322.1348M} were not only able to image 3 exoplanets associated with HR 8799, but also detect their orbital motions throughout high spatial resolution imaging spread throughout a time-span of 4 years. The beauty of direct imaging exoplanets is that, with the high spatial resolution delivered by AO, we can even separate exoplanet's photons from that emitted by the host star, and feed exoplanet's light into spectrograph to study the exoplanetary atmospheres directly. There are several disadvantages of direct imaging, though. The first one is since we need AO to separate the exoplanets from the stars, this means we are limited to only bright (or nearby) stars where AO can perform properly to compensate the atmospheric turbulences. The other drawback of direct imaging is that, even though we can use coronagraph to block the light from the star and achieve high contrast imaging, we still can only probe exoplanets at larger orbital distance where we can remove the host star's light contribution properly. Indeed, most of the exoplanets discovered by the direct imaging method are very far away from their host stars (tens to hundreds of AU), posing challenges to \textit{in situ} planet formation scenarios via either core accretion or disk instability. Currently there are two proposed formation mechanisms for such wide separation exoplanets. The first one is the widely separated planets and the host stars are formed like binary or multiply stellar system via cloud fragmentation in the natal environment \cite{2009ApJ...694..183F}. Another possibility is that these wide separation exoplanets are formed by planet-planet scattering \cite{1996Sci...274..954R}, where there were two exoplanets formed via core accretion in the natal proto-planetary disk. However, the orbits of these two exoplanets are not stable, and through dynamical interactions, one exoplanet was scattered to a very large distance, while the other one migrated inwards and formed a very close-in orbit. Unfortunately the true formation mechanism of wide-separation exoplanets are still under debate. Detections of multiple exoplanet systems -- with a very close-in exoplanet and a wide-separation exoplanet orbiting the same host star -- would provide a smoking gun evidence to support such scenario.  

\section{Status quo}
With the successful progresses of various exoplanet detection methods, especially with the discovery of more than 3,000 exoplanets, we are now gathering a large sample of exoplanets that enables us to perform statistical studies and answer key questions about planet formation and evolution.  

To address these questions, we first need to have precise and accurate measurements of the exoplanetary parameters, especially their masses and radii. For this, we need to join the forces of both transit and radial velocity methods. The former provides a good estimate of the radius of the exoplanets assuming that we know the host star's radius. This is because from transit light curve, we can infer the radius ratio between the exoplanet and the host star via the depth of the transit dipping. More importantly, from the transit light curves, we can determine the inclination of the exoplanet orbit, and therefore break the degeneracy of the unknown inclination angle issue that are prevailing in the radial velocity method. With the large sample of transiting exoplanets discovered by Kepler, we are now having radius estimate for thousands of exoplanets. However, to really address questions in planet formation and evolution, we need to have a good control on the error of the radius estimates, which mainly comes from the uncertainties in the host star's properties. In this regard, the California Kepler Survey, with dedicated high spectral resolution follow-up of Kepler planet host stars using the High Resolution Echelle Spectrometer (HIRES) instrument mounted on the Keck telescope, is delivering high precision estimates of $\sim$ 1,000 Kepler exoplanet host stars \cite{2017AJ....154..107P}, and in turn provides precise and accurate estimate of the radius of $\sim$ 2,000 exoplanets \cite{2017AJ....154..108J}. With this unique sample where the exoplanetary properties are determined with high precision, Fulton et al. \cite{2017AJ....154..109F} were able to investigate the radius distritubion of exoplanets in detail, where they found that there are two distinct populations of exoplanets, i.e. super-Earths with radii peaked at $\sim$ 1.3 R$_{Earth}$ and sub-Neptunes with radii peaked at $\sim$ 2.4 R$_{Earth}$. In addition, there seems to be a dearth of exoplanets with radii $\sim$ 1.8 M$_{Earth}$, hinting to a evaporation valley at this radius. 

With the high precision radial velocity in hand, we can also determine the mass of the exoplanets. In this regard, there are hundreds of precise and accurate mass estimates \cite{2017ApJ...834...17C}. The precise mass estimates, along with the radius estimates, can provide us insights to the mass-radius relation of the exoplanets. Chen \& Kipping \cite{2017ApJ...834...17C} investigated the exoplanetary mass-radius relations in detail with a sample of $\sim$ exoplanets, resulting in probabilistic forecasts of piece-wise mass-radius relations across terrestrial, Neptunian, and Jovian masses, where there is a transition at $\sim$ 2 M$_{Earth}$, dividing the terrestrial rocky planets from Neptunian exoplanets, as well as a transition at $\sim$ 0.41 M$_{Jupiter}$, dividing the Neptunian exoplanets from Jovian exoplanets. While the mass-radius relation can tell us on the mean density of exoplanets, there is still a degeneracy in the exoplanet composition, i.e. whether it is an exoplanet with small rocky core and large water-rich envelope, or an exoplanet with a relative large rocky core with thiner Hydrogen/Helium envelope.  

For this we will need to study the atmospheric composition of the exoplanets directly. While the light emits by the exoplanets themselves are very faint, it is still possible to study their atmosphere with transmission spectroscopy. During the the transit, the exoplanetary atmosphere are back-lit by the host star, so we can study the exoplanetary atmospheric properties by comparing the host star spectra during and outside of the transit. However, even with the transmission spectroscopy, the exoplanetary signature is very faint so most of the work are done with Jovian exoplanets. In addition, the transmission spectroscopy can be very vulnerable to seeing variation and slit loss if conducted from the ground. In this regard, most of the transmission spectroscopy are either carried out from the space using Hubble Space Telescope or Spitzer, or carried out in a relative manner -- i.e. using comparison stars with similar spectral type within the field of view -- from ground-based telescope. Indeed, Sing et al \cite{2016Natur.529...59S}. have used Hubble Space Telescope and Spitzer to obtain transmission spectra of 10 hot Jupiters covering the wavelength of 0.3 to 5 $\mu$m. Their results suggest that exoplanetary atmosphere of even hot Jupiters can come into different flavors, with clear to cloudy/hazy.

Another basic parameter of exoplanets is their orbital distance.
It is true that the first few exoplanets discovered are very close to their host stars (so-called hot Jupiters) due to limitations of techniques employed at earlier days. However, as instruments were upgraded with higher precision, more and more exoplanets are found at larger distance, temperate to have liquid water and habitable for lifes \citep{2013Sci...340..587B,2016Natur.533..221G}. For example, Kepler has found almost 30 exoplanets in the habitable zone\footnote{https://www.nasa.gov/kepler/discoveries}.
In addition, from the orbital distance vs. mass distribution, we can have a glimpse of their formation process. There have been two proposed \textit{in situ} scenarios for planets formed in the protoplanetary disks. The first one is core accretion \cite{1996Icar..124...62P}, where small particles collide and coagulate into planet embryos that later on accrete gases from the surrounding environment and form Jovian planets. However, there are some massive exoplanets at larger orbits which cannot be explained by core accretion along, as the accretion is not efficient to form such planets. In this regard, the second planet formation mechanism is disk instability \cite{2000ApJ...536L.101B}, where as the protoplanetary disks cool down, there are turbulences lead to high density in some regions and form self-gravitating gas clumps. These clumps then trigger rapid accumulation of gases from the surrounding regions and formed gaseous planets. 

One observational evidence to test the aforementioned planetary formation scenarios is via metallicity (the fraction of mass of a star that is not H or He). This is because the core accretion is more efficient in metal rich environment, while the disk instability has no metallicity dependence. In this regard, there have been investigations of the metallicity of exoplanetary host stars. For example, Fisher \& Valenti \cite{2005ApJ...622.1102F} showed that gaseous are more common around metal rich stars. In addition, studies by Thorngren et al. \cite{2016ApJ...831...64T} also showed that gaseous giant exoplanets have enhanced metallicity, especially significant amounts of heavy elements in their H/He envelopes. These piece of evidences support that most of the exoplanets are formed via core accretion.

When studying exoplanets, we should also keep in mind that exoplanets migrate, and they may not form at their present day location. This can be seen from studies of planets in our solar system \citep{2005Natur.435..466G,2012Natur.485...78B}. There are also indirect evidences from exoplanets, for example, the direct imaging method has discovered a dozen of exoplanets that are tens to hundreds AU away from their host stars \cite{2016PASP..128j2001B}. These wide-separation exoplanets are difficult to be explained by the aforementioned \textit{in situ} formation processes. One possibility is that they are formed by planetary scattering \cite{1996Sci...274..954R}, where there are at least two exoplanets formed in the protoplanetary disk, but their orbits are not stable and cross each other. Due to this dynamical interaction, one exoplanet was ejected to a very wide orbit, while the other exoplanet migrated in to be a very close-in exoplanet. Another alternative theory is that the widely separated exoplanet is not really orbiting the host star, but rather a low mass companion formed in the same molecular cloud similar to the process of binary / multiple star formation, which is often dubbed as the cloud fragmentation scenario \cite{2009ApJ...694..183F}. 

\begin{figure}[!t]
\centering
\includegraphics[width=16cm]{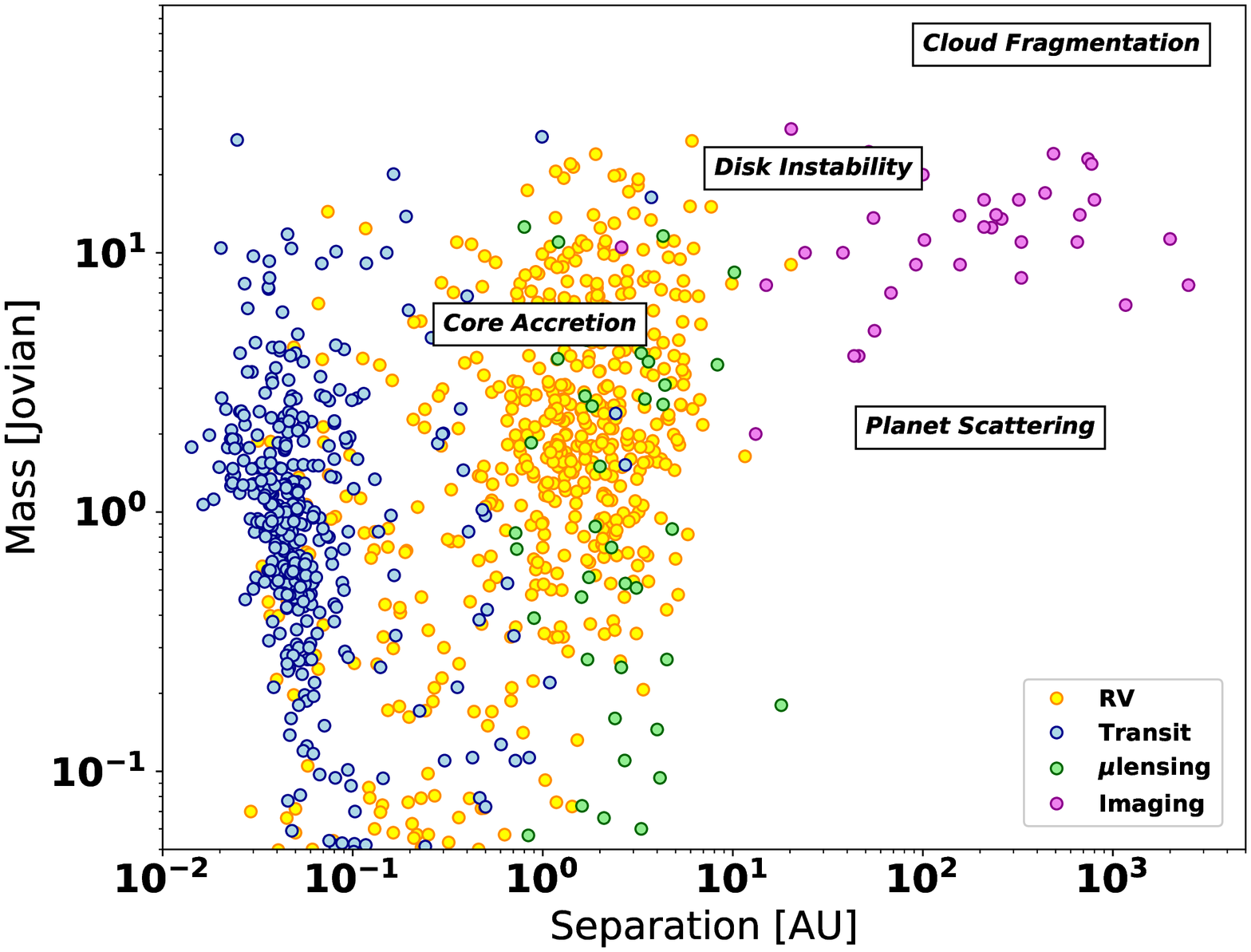}
\caption{Orbital distances and mass distributions of exoplanets from radial velocity, transit, microlensing, and direct imaging methods. Data gathered from NASA Exoplanet Archive.}
\end{figure}   

There are several ways to distinguish the planetary scattering from cloud fragmentation scenarios \cite{2016PASP..128j2001B}. For example, the planet-planet scattering scenario will produce highly eccentric exoplanets, while the exoplanets formed via cloud fragmentation will have eccentricity distribution similar to binary stars. If there are multiple exoplanets in a system, the architecture of these exoplanetary system can also shed lights on their formation process. For example, wide-separation exoplanets formed by planet-planet scattering with have orbits that are coplanar with the protoplanetary/debris disk. Wide-separation exoplanets that are formed by cloud fragmentation, on the other hand, will have orbits that are not coplanar with the protoplanetary/debris disk, nor with the stellar spin axis. The formation timescale (or age of the planetary system) can also be used as an indication of the formation process. In general, cloud fragmentation will produce planetary mass objects within one million years; the planet-planet scattering mechanism, on the other hand, will form gaseous ginat planets between one to ten million years. Future observations in these directions will give us insight into the planet formation mechanism for the widely separated exoplanets.

Beyond widely-separated exoplanets, astronomers are also interested in free floating (or unbound) exoplanets. This is especially the case not only to understand their formation process (can exoplanets form in an isolate manner like brown dwarfs?), but to characterize the mass function at the very low end. However, these free floating exoplanets are difficult to capture, especially they are too faint to be identified even with the state-of-the-art large telescopes on the ground. In this regard, microlensing provides the best way to identify them, especially because microlensing only relies on the gravitational field exerted by the planets, and hardly depends on their electro-magnetic emissions. The only hurdle for microlensing is that the event time-scale is proportional to the mass of the underlying lens, and as the lens mass goes as light as Jovian planets, the time-scale is in the order or merely 1 day (or even less) and requires continuously monitoring with intra-night cadence. To overcome this hurdle, the MOA team has conducted a very high cadence (every 10 to 50 minutes) survey and discovered a dozen of candidate free-floating exoplanets \cite{2011Natur.473..349S}. Their result also suggests that a wealth of Jovian mass objects, as twice populous as main-sequence stars, exists in the Milky Way. This causes an abrupt change in the mass function below brown dwarf masses, and suggests that the free-floating exoplanets are formed differently, i.e. they might first formed in the protoplanetary disks, and subsequently ejected to unbound or at very distant orbit. However, such wealth of Jovian mass objects is hard to be reconciled by ejection mechanism \cite{2012MNRAS.421L.117V}. In this regard, the OGLE team repeated the same investigations using a much larger sample from OGLE-IV \cite{2017Natur.548..183M}, and they did not find such excess of Jovian mass objects. However, they found some candidate ultra-short events ($<$ 0.5 days), indicating the presence of free-floating terrestrial or super-Earth mass exoplanets, which is in good agreement with the predictions from core-accretion and planet-planet scattering mechanisms.   

Another important aspect of our understanding of exoplanets is how do exoplanets form and evolve with time. In this regard, direct imaging method may provide the most comprehensive picture, from the birth of exoplanets in the protoplanetary disk, till the disk completely dissipated and only left with debris disk. This is only possible by combining the direct imaging power from both of the near infrared coronagraph coupled with AO at large optical telescopes, and the exquisite spatial resolution by millimeter arrays. For example, ALMA observations of HL Tau, a $\sim$ 1 Myr T Tauri star, provides a detail view of protoplanetary disk at the stage where the exoplanets are about to be born \cite{2015ApJ...808L...3A}. On the other hand, VLT/NACO observation of $\beta$ Pic -- a $\sim$ 10 Myr solar mass star -- provides the first view of both exoplanets and circumstellar disks \cite{2010Sci...329...57L}. Future direct imaging observations will provide us insight not only in the formation and evolution of exoplanets, but their co-evolution with protoplanetary disks as well.
 
%%%%%%%%%%%%%%%%%%%%%%%%%%%%%%%%%%%%%%%%%%
\section{New Frontiers}

Advances in exoplanetary studies come hand-in-hand with new instruments and surveys. This is especially the case for the radial velocity method. For example, the Earth will exert a radial velocity signal of merely $\sim$ 9 cm/s. This is beyond the reach of the state-of-the-art high resolution spectrograph, e.g. HARPS. To achieve this goal, there have been constructions of next generation instruments, both at the VLT and the upcoming E-ELT. For example, the recently commissioned ESPRESSO \cite{2010SPIE.7735E..0FP} at VLT will be able to reach $\sim$ 10 cm/s precision. The CODEX instrument planned for E-ELT is aiming for $\sim$ 2 cm/s precision \cite{2008SPIE.7014E..1IP}. 
 
However, the main hurdle of precision radial velocity measurements is not from instrumentation, but actually from stellar variations, e.g. stellar spots, magnetic activities, that can generate radial velocity signals as much as several m/s and overshadow the radial velocity variations from exoplanets, especially terrestrial ones. One example is HD 154345, which showed radial velocity variations and was first thought to be caused by a companion exoplanets \cite{2008ApJ...683L..63W}, but later  turned out to be from stellar activity instead \cite{2016arXiv160308384W} instead. In this regard, we need to monitor the stars both photometrically and spectroscopically, where the photometric observations can help us sort out the stellar variabilities' influence on the radial velocity measurements \cite{2017A&A...606A.107O}.

There has been a frog leap in the number of transiting exoplanets thanks to the continuous monitoring and high precision photometry delivered by the Kepler space telescope. While Kepler is not longer observing its originally scheduled field due to the malfunction of two gyroscopes, it is now transforming into the K2 mission and still delivering many new transiting exoplanets despite the selected fields. There will be Transiting Exoplanet Survey Satellite (TESS\footnote{https://tess.gsfc.nasa.gov/}) mission in the immediate future and the Planetary Transits and Oscillations of stars (PLATO\footnote{http://sci.esa.int/plato/}) mission is also planned in the long run, both are going to monitor bright stars from space, with the hope to find much tiny transit signals. The beauty of transit method is that it can investigate the exoplanetary atmosphere back-lit by the stars. In this regard, ESA recently announced the selection of ARIEL, the Atmospheric Remote-sensing Infrared Exoplanet Large-survey mission, as the next medium class mission to be launched in 2028\footnote{http://sci.esa.int/ariel/}. The transit searches can also be carried out from the ground, preferentially smaller apertures to enable large field-of-view and survey the entire sky in a fast manner. For example the Next Generation Transit Survey (NGTS) \cite{2016Msngr.165...10W}, the Kilodegree Extreme Little Telescope (KELT) survey \cite{2007PASP..119..923P}, the Transiting Planets and Planetesimals Small Telescope (TRAPPIST) survey \cite{2011EPJWC..1106002G}, the Exoplanets in Transits and their Atmospheres (ExTrA) survey \cite{2015SPIE.9605E..1LB}, to name a few, are all patrolling the entire sky or being built from all around the globe. 

One caveat of the current transit surveys it that they sacrifice spatial resolution to obtain large field-of-view, meaning that we will need to sort out blending effect. This is especially important because if a unrelated background star (or galaxy) also contributes to the same pixel of the CCD, this will cause misinterpretation of the exoplanet's size if we assume the photons only come from one single (host) star. For example, in a blended case we might infer the exoplanet as terrestrial size, but after deblending, it will actually be of Jovian size \cite{2015ApJ...804...97C}. In this regard, it is important to have high spatial resolution imaging to gauge the blending influence on the host stars. While it can be difficult and inefficient for canonical AO observation, where AO optimization takes lots of overheads, the Robo-AO team has demonstrated that with a robotic manner, it is possible to tackle the deblending issue with automated AO system, even to deliver a sample of thousands of host stars in a comparatively short time scale (few years) \cite{2017arXiv171204454Z}.
 
%%%%%
Similar to transit method, the microlensing method will also see a boom in the number of detected events with the arrival of the future space-born mission -- WFIRST. Continuous monitoring of the Galactic bulge by WFIRST from the space, especially with its wide field-of-view and exquisite spatial resolution, will provide high cadence, high quality light curves that are expected to deliver discoveries of thousands of exoplanets, and can even enable us to detect exoplanets with masses as low as Jovian satellites. In addition, concurrent observation from both the space and ground will provide the parallax information that can help us break the microlensing degeneracy and pin down the ensemble mass of the lens system -- i.e. the combination of the host star and the exoplanet. We need to keep in mind that from the microlensing light curve we can only derive the mass ratio between the host star and the exoplanet, and we need further pieces of information to constrain the mass of the host star. In this regard, the exquisite spatial resolution can also help us to disentangle the lens light from the source several years after the peak of the microlensing event, where the photometry of the lens can be used to pin down the mass of the host star and the exoplanet \cite{2017Univ....3...53L}.

As mentioned before, it is important to have concurrent observations from both the space and the ground to break degeneracy in microlensing. In this regrad, there have already been dedicated surveys, e.g. OGLE\footnote{http://ogle.astrouw.edu.pl/}, MOA\footnote{http://www.phys.canterbury.ac.nz/moa/}, KMTNet\footnote{http://kmtnet.kasi.re.kr/kmtnet-eng/}, each surveying the Galactic bulge with very high cadence and wide field of view. In the future, there is also going to be the PRime-focus Infrared Microlensing Experiment (PRIME) survey, using the prime focus at the 1.8m telescope at the SAAO, enabling a wide-field view with a dedicated infrared camera (FOV $\sim$ 1.56 deg$^2$, the largest in the world). The benefit of going into the infrared is that we are less susceptible to the extinction effect, which is especially sever toward the Galactic center, and can penetrate better into the Galactic bulge for microlensing events. In addition, the Large Synoptic Survey Telescope (LSST\footnote{https://www.lsst.org/}) is also going to have high cadence survey of the entire southern sky, delivering exquisite light curves in the optical with five different filters, providing complementary color information of the microlensing events.   

The direct imaging method is also expected to be transformed by the arrival of new extreme AO systems and large surveys. For example, the recently commissioned Gemini Planet Imager (GPI) \cite{2014SPIE.9148E..0JM} is conducting a comprehensive exoplanet surveys (GPIES\footnote{http://www.gemini.edu/science/GPI/proposal\_macintoshv2.pdf}) of 600 stars with spectral type ranging from A to M, aiming not only to detect exoplanets, but debris disk as well, to understand the formation and architecture of gaseous exoplanets at 5-50 AU. New techniques, e.g. the Phase Induced Amplitude Apodization (PIAA) Coronagraph \cite{2005ApJ...622..744G} on-board implemented in the Subaru Coronagraphic Extreme Adaptive Optics (SCExAO) \cite{2015PASP..127..890J}, will deliver very high contrast capability at very small separation \cite{2017ApJ...836L..15C}. The planned starshade spacecraft \cite{2015SPIE.9605E..0WS} -- also known as the external occulter -- is aiming at a contrast level of 10$^{-10}$ and will enable WFIRST to directly imaging Earth-like exoplanets in habitable zone. 

With the aforementioned surveys/technologies, we will not only be able to study the formation and evolution of exoplanets, but to learn if there are other pale blue dots in the Universe.

%%%%%%%%%%%%%%%%%%%%%%%%%%%%%%%%%%%%%%%%%%
\vspace{6pt} 

%%%%%%%%%%%%%%%%%%%%%%%%%%%%%%%%%%%%%%%%%%
%% optional
%%%%%%%%%%%%%%%%%%%%%%%%%%%%%%%%%%%%%%%%%%
\acknowledgments{The author is indebeted to the referees, whose comments greatly improved the manuscript.}

%%%%%%%%%%%%%%%%%%%%%%%%%%%%%%%%%%%%%%%%%%
%\authorcontributions{For research articles with several authors, a short paragraph specifying their individual contributions must be provided. The following statements should be used ``X.X. and Y.Y. conceived and designed the experiments; X.X. performed the experiments; X.X. and Y.Y. analyzed the data; W.W. contributed reagents/materials/analysis tools; Y.Y. wrote the paper.'' Authorship must be limited to those who have contributed substantially to the work reported.}

%%%%%%%%%%%%%%%%%%%%%%%%%%%%%%%%%%%%%%%%%%
\conflictofinterests{The author declare no conflict of interest.} 

%%%%%%%%%%%%%%%%%%%%%%%%%%%%%%%%%%%%%%%%%%
%% optional
%\abbreviations{The following abbreviations are used in this manuscript:\\

%\noindent MDPI: Multidisciplinary Digital Publishing Institute\\
%DOAJ: Directory of open access journals\\
%TLA: Three letter acronym\\
%LD: linear dichroism}

%%%%%%%%%%%%%%%%%%%%%%%%%%%%%%%%%%%%%%%%%%
%% optional

%%%%%%%%%%%%%%%%%%%%%%%%%%%%%%%%%%%%%%%%%%
\bibliographystyle{mdpi}

%=====================================
% References, variant A: internal bibliography
%=====================================
\renewcommand\bibname{References}

%=====================================
% References, variant B: external bibliography
%=====================================
%\bibliography{your_external_BibTeX_file}

%%%%%%%%%%%%%%%%%%%%%%%%%%%%%%%%%%%%%%%%%%
%% optional
%\sampleavailability{Samples of the compounds ...... are available from the authors.}

%%%%%%%%%%%%%%%%%%%%%%%%%%%%%%%%%%%%%%%%%%
\end{document}